# Effects of Pulse Duration on Kerr Self-Focusing and Weld Formation in Ultrafast Laser Welding of Fused Silica

YITONG CHEN,[1] NAN LI,[2,4] ZHIJIE HAN,[2] YUXUAN LI,[2,3] QINGWEI ZHANG,[2] ZIHUI DONG,[2] YUMING SUN,[2] ZHE LIN,[2] GANGYI ZHU,[1,5] SHANGLU YANG[2,3,*]

[1]National and Local Joint Engineering Laboratory of RF Integration and Micro-assembly Technology, College of Integrated Circuit Science and Engineering, Nanjing University of Posts and Telecommunications, Nanjing, 210003, China
[2]Shanghai Institute of Optics and Fine Mechanics, Chinese Academy of Sciences, Shanghai 201800, China
[3]Center of Materials Science and Optoelectronics Engineering, University of Chinese Academy of Sciences, Beijing 100049, China
[4]linan@siom.ac.cn
[5]zhugangyi@njupt.edu.cn
*yangshanglu@siom.ac.cn

**Kerr self-focusing can shift the laser interaction region away from the geometrical focus, making energy deposition near the joining interface difficult to control. Here, the effect of pulse duration on nonlinear focusing and weld formation is investigated from 262 fs to 10 ps. ICCD imaging shows clear changes in the axial position and shape of the plasma emission. Different pulse durations already produce distinct axial positions during the initial 2000 pulses, and these positions continue to evolve with pulse accumulation. The final weld morphology, weld position, and shear strength show corresponding changes. A ray-optics model including the Kerr effect is used to calculate the focal shift. The simulated trend agrees with the experiment and supports the role of Kerr self-focusing in the observed focal position change. These results clarify the relationship between pulse duration, nonlinear focusing, interaction position, and weld formation.**

Fused silica is widely used in precision optics because of its broad transparency, chemical stability, and low thermal expansion. Its use is also expanding in compact optoelectronic and implantable systems. Implantable neurostimulation platforms, for example, have used fused silica enclosures sealed by femtosecond laser welding [1]. Focused ultrashort pulses can deposit energy locally inside transparent glass through nonlinear absorption [2-4]. At high repetition rates, successive pulses produce heat accumulation and local melting near the interface, enabling direct joining [5-9]. Previous studies have demonstrated glass welding under different interfacial conditions, including crack-free welding, gap bridging, and welding without optical contact [10-13].

The spatial position of energy deposition is critical to weld formation. In transparent dielectrics, nonlinear propagation can shift the interaction region away from the geometrical focus. At fixed pulse energy, reducing the pulse duration increases the peak power and peak intensity. Kerr self-focusing can shift the high intensity region toward the incident side, whereas plasma generation can partly counteract this focusing [14, 15]. In fused silica, nonlinear propagation has also been shown to depend strongly on pulse duration [16]. The critical power for self-focusing changes from the sub-picosecond to picosecond regime [17]. In laser welding, focal pre-compensation has been used to improve energy deposition near the interface [18]. However, how the nonlinear focusing position changes with pulse duration during pulse accumulation, and how this change affects weld formation, remain unclear.

Here, we investigate fused silica welding from 262 fs to 10 ps with the geometrical focus fixed 1.4 mm below the interface. The final weld morphology, weld position, and shear strength are first compared. ICCD imaging is then used to record the plasma emission over selected 2k-pulse intervals during fixed point irradiation. Additional ICCD measurements during scanning welding are used to compare the interaction position at different pulse durations. A COMSOL ray-optics model including the Kerr effect is used to evaluate the change in focal position. The measurements and simulation are used to examine the relationship between pulse duration, nonlinear focusing, interaction position, and weld formation.

Two fused silica samples with dimensions of $10 \times 10 \times 3\ mm^3$ were stacked for welding. Figure 1 shows the experimental setup. A 1030-nm ultrafast laser was operated at 100 kHz, with the pulse duration varied from 262 fs to 10 ps. The scanning speed and focal spot diameter were 100 mm/s and 40 µm, respectively. The average power was fixed at 16 W for all pulse durations. Thus, the pulse energy remained unchanged, while the peak power and peak intensity varied with pulse duration. After beam expansion, the beam was directed by a galvo scanner and focused into the sample using an $f = 70$ mm F-theta lens. The geometrical focus was fixed 1.4 mm below the interface for all pulse durations. Parallel scan lines, 5 mm in length and 300 µm apart, were used to form the welded area [Fig. 1(b)]. Plasma emission was recorded from the side using an ICCD camera, with the viewing direction perpendicular to the laser propagation direction. For the pulse accumulation measurements, fixed point irradiation was used with a total exposure of 10k pulses. The ICCD exposure time corresponded to 2k pulses, and its trigger delay was adjusted using a DG645 digital delay generator to record selected intervals from 0–2k to 8–10k pulses. Each interval was measured independently at a fresh sample position.

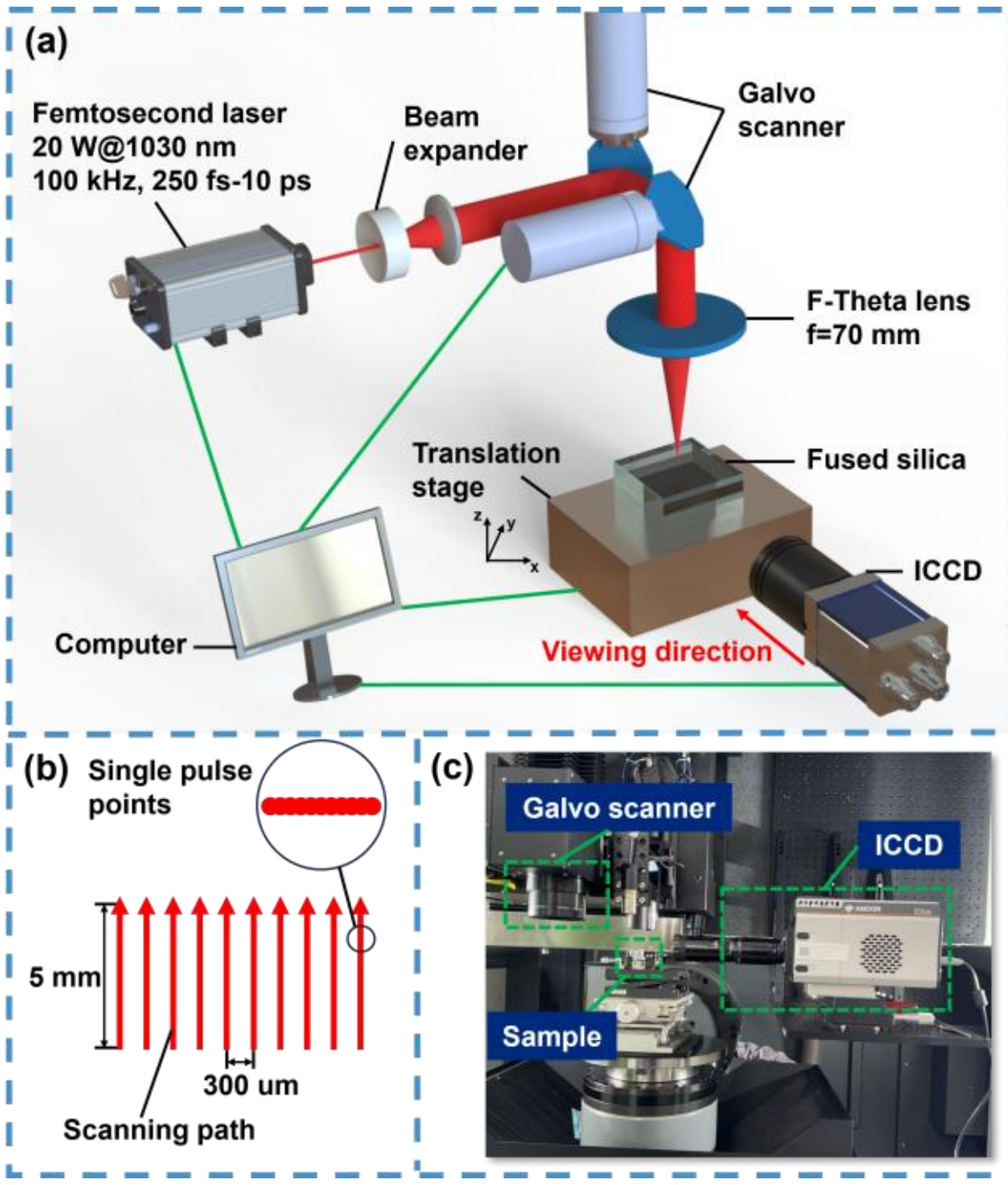

Fig. 1. Experimental setup for ultrafast laser welding of fused silica. (a) Schematic of the laser welding and ICCD imaging system. (b) Laser scanning path on the sample. (c) Photograph of the experimental setup.

The final weld characteristics were first evaluated over the full pulse duration range, as shown in Fig. 2. The modified zone broadened as the pulse duration increased, with $d$increasing from 34.0 μm at 262 fs to a maximum of 89.4 μm at 6 ps. The length $\Delta L$ increased rapidly in the femtosecond range and changed only moderately at longer pulse durations. A much larger variation was observed in the weld position. Δz decreased from 1503 μm at 262 fs to −333 μm at 10 ps, indicating a pronounced downward displacement of the weld region with increasing pulse duration. In contrast, the mean shear strength generally decreased with pulse duration, with a maximum value of 76.1 MPa at 262 fs.

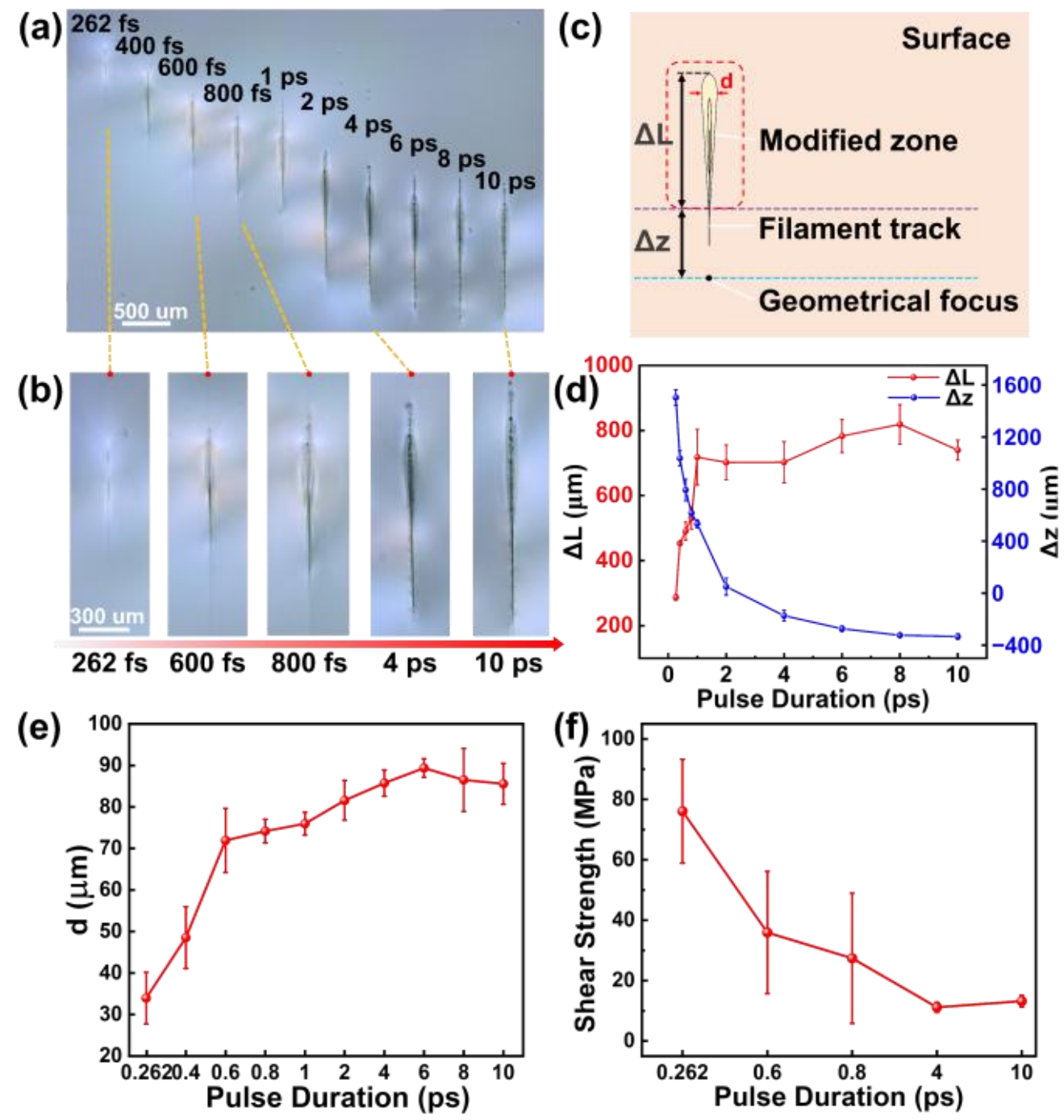

Fig. 2. Weld morphology and shear strength at different pulse durations. (a) Overall weld morphologies obtained from 262 fs to 10 ps. (b) Enlarged views of representative welds. (c) Definition of the modified zone, filament track, geometrical focus, the width of modified zone $\boldsymbol{d}$, the length of modified zone $\boldsymbol{\Delta L}$, and axial offset $\boldsymbol{\Delta z}$. (d) $\boldsymbol{\Delta L}$ and $\boldsymbol{\Delta z}$, (e) $\boldsymbol{d}$, and (f) shear strength as functions of pulse duration.

Figure 3(a) shows the plasma emission recorded over these pulse intervals. The emission evolves differently with pulse accumulation at different pulse durations. At 262 fs, the emission region becomes progressively longer, whereas its end remains nearly fixed at about 1.38 mm above the geometrical focus [Fig. 3(b)]. At 800 fs, the emission length changes less, but the emission end moves upward from 491 to 844 μm above the geometrical focus. A different behavior appears in the picosecond range. At 2 ps, the emission end starts below the geometrical focus and moves toward, and later slightly above, it with pulse accumulation. At 6 and 10 ps, the emission region shortens strongly after the first 2k pulses, while its end moves from below to above the geometrical focus. At 6 and 10 ps, the emission became too weak at later pulse intervals to define a reliable axial position.

Figure 3(c) isolates the first 0–2k pulses and shows that the difference is already present at the early stage of irradiation. The emission end moves downward with increasing pulse duration, from 1377 μm above the geometrical focus at 262 fs to 491 μm above it at 800 fs. At 2 ps, it is already 263 μm below the geometrical focus and moves farther downward to 412 μm below it at 10 ps. The largest positional change therefore occurs from the femtosecond to the early picosecond range. These results show that pulse duration dependent differences in the emission position are already present within the first 2k pulses, while subsequent pulse accumulation further changes the emission position and length.

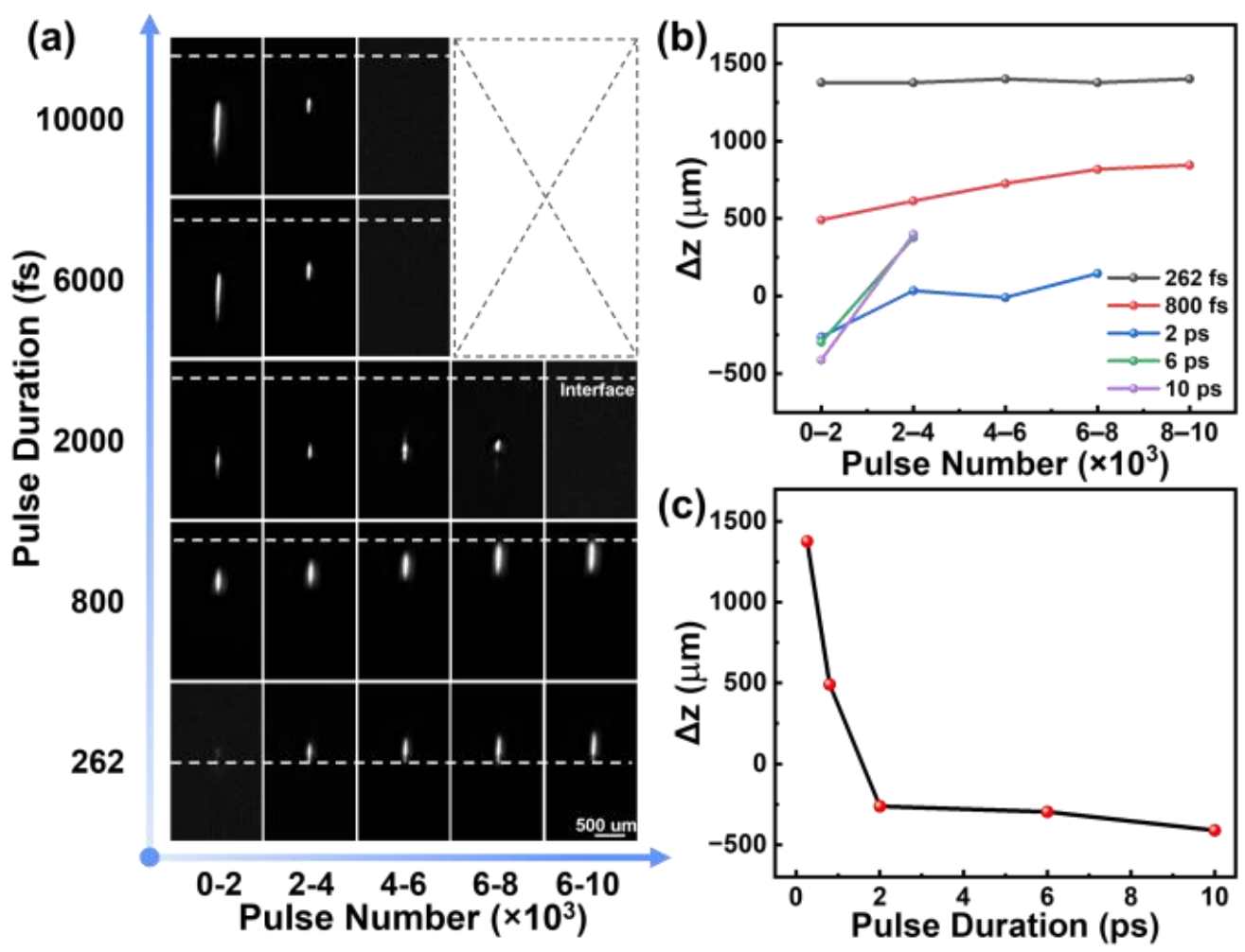

Fig. 3. Evolution of plasma luminescence during pulse accumulation at different pulse durations. (a) ICCD images acquired over selected 2k-pulse intervals during a total exposure of 10k pulses. (b) Axial offset from the end of the plasma luminescence to the geometrical focus during pulse accumulation. (c) Axial offset obtained during the initial 2k-pulse interval as a function of pulse duration.

The pulse duration dependence of the interaction region was further examined during scanning welding. As shown in Fig. 4(a), shorter pulses produced an emission pattern with multiple bright maxima and an elongated filamentary region. With increasing pulse duration, the emission region shifted downward toward the geometrical focus and became more localized. The position of maximum plasma luminescence was used as an experimental indicator of the effective nonlinear focusing position. The characteristic regions are defined schematically in Fig. 4(b), where Δh denotes the axial distance between the effective nonlinear focus and the geometrical focus.

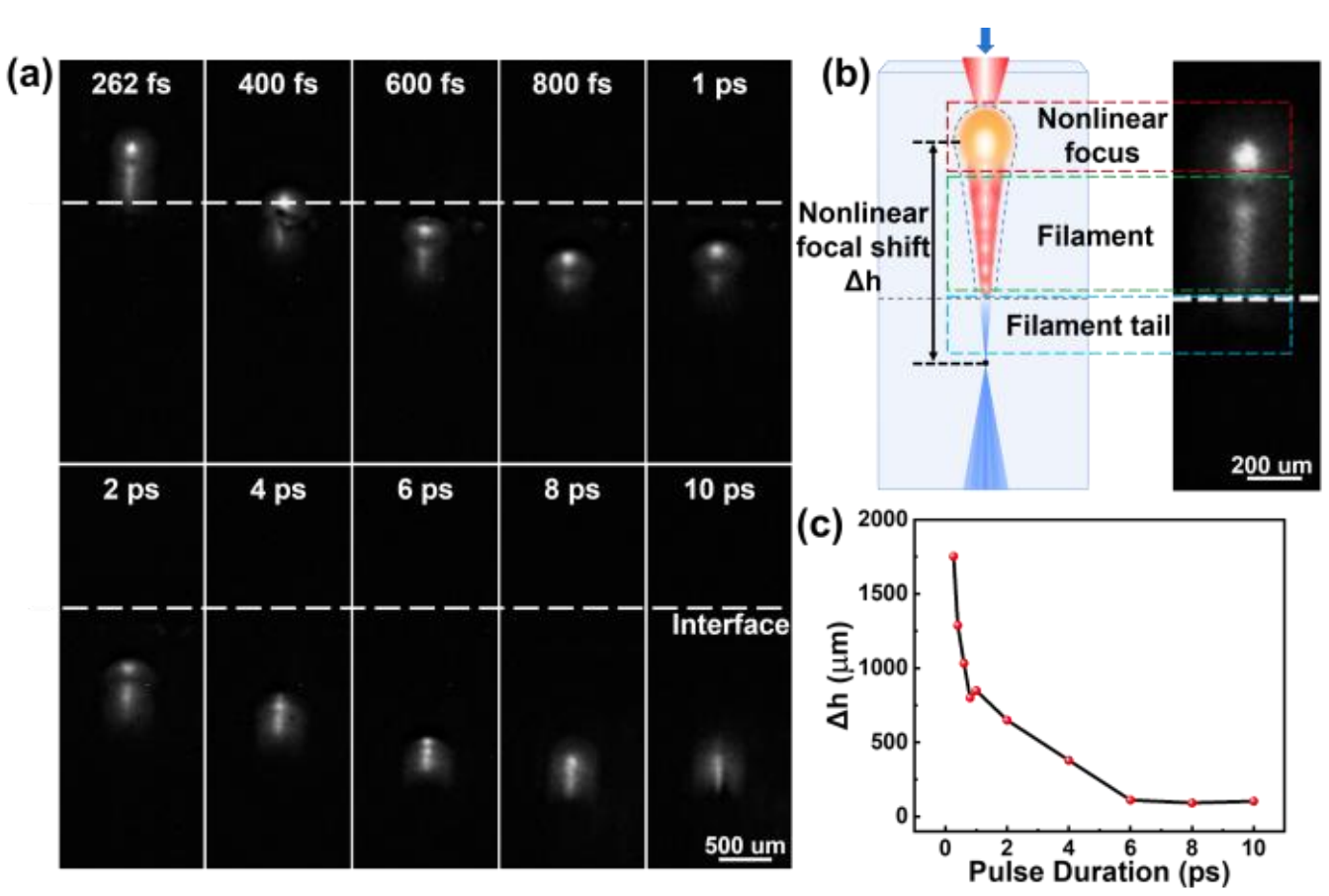

Fig. 4. Plasma luminescence and nonlinear focal shift at different pulse durations. (a) ICCD images of plasma luminescence during scanning welding from 262 fs to 10 ps. (b) Schematic definition of the nonlinear focus, filament, filament tail, and nonlinear focal shift $\boldsymbol{\Delta h}$. (c) Nonlinear focal shift $\boldsymbol{\Delta h}$ as a function of pulse duration.

Figure 4(c) shows a clear dependence of Δh on pulse duration. It decreases from 1753 μm at 262 fs to about 800 μm at 800 fs, and then to 650 μm at 2 ps. The decrease becomes much slower at longer pulse durations. At 6 ps, Δh is about 111 μm and remains close to 100 μm from 8 to 10 ps. The largest change therefore occurs from the femtosecond to the early picosecond range, whereas the effective nonlinear focusing position varies only slightly at longer pulse durations.

To examine the origin of the observed focal shift, the Kerr effect was included in the COMSOL ray-optics model using $n = n_0 + n_2 I$, where $n_0$ is the linear refractive index, $n_2$ is the nonlinear refractive index coefficient, and I is the laser intensity. The linear refractive index and nonlinear refractive index coefficient of fused silica were taken from established optical-property data [19, 20]. Figure 5(a) shows the calculated axial distribution of the induced refractive index change. At the same pulse energy, shorter pulses have a higher peak intensity and produce a larger Kerr-induced refractive index change. This corresponds to stronger nonlinear focusing. The calculated focal shift follows the experimental trend [Fig. 5(b)]. Both decrease rapidly with increasing pulse duration and approach smaller values in the picosecond range, although the simulated shifts are generally larger than the measured values. This agreement in trend supports the contribution of Kerr self-focusing to the observed focal shift.

The corresponding focusing behavior and weld formation are illustrated in Figs. 5(c) and 5(d). In the femtosecond regime, stronger Kerr self-focusing shifts the effective focus farther toward the incident side, away from the geometrical focus. As the pulse duration increases, the nonlinear focusing becomes weaker, and the effective focus moves closer to the geometrical focus. This difference is consistent with the weld position variation observed in Fig. 2, where shorter pulses produce narrower modified zones at higher axial positions, while longer pulses form broader modified zones closer to the geometrical focus. Although the model does not describe the complete welding process, the results support the role of Kerr self-focusing in determining the interaction position and subsequent weld formation.

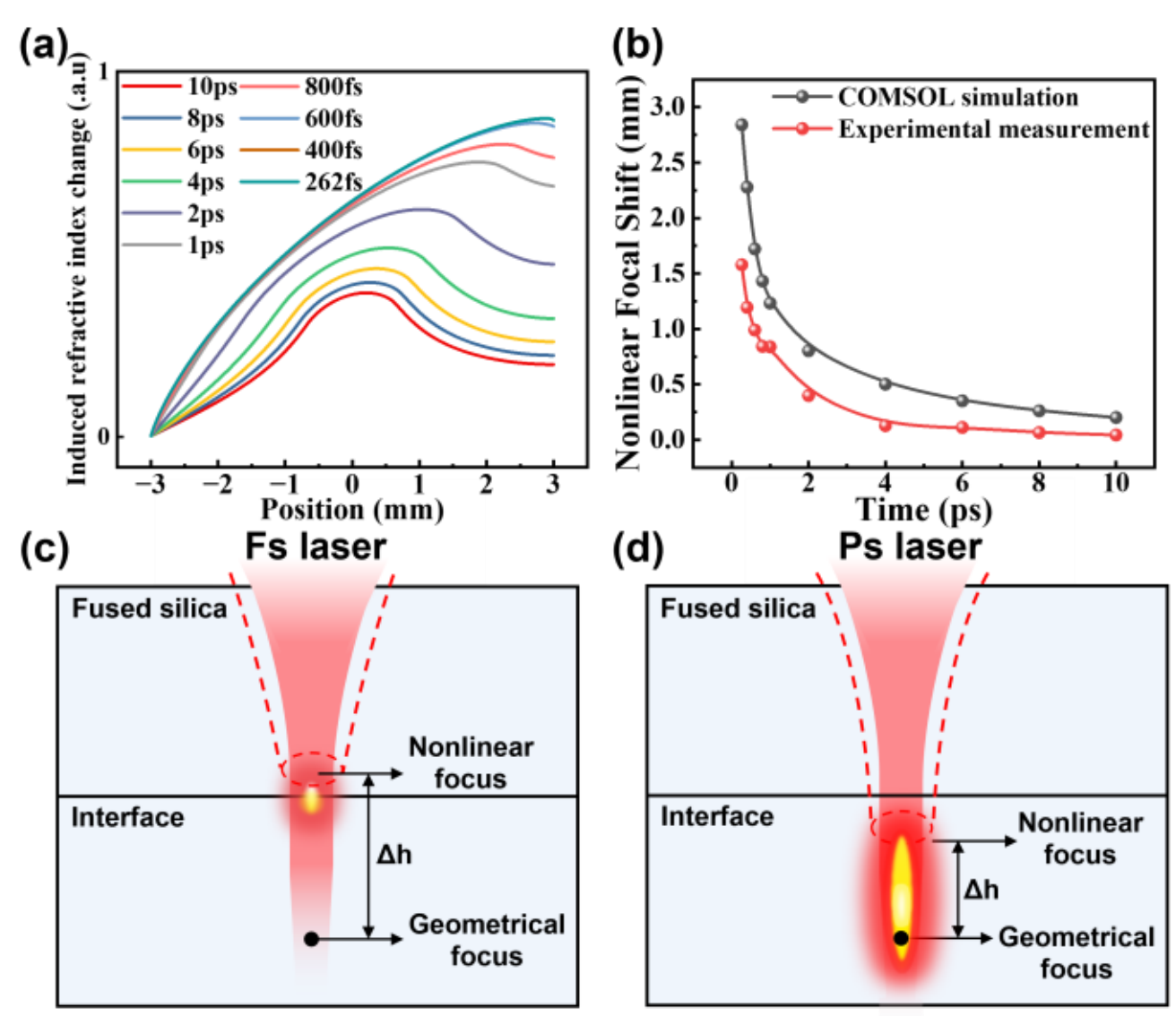

Fig. 5. COMSOL simulation and schematic interpretation of pulse duration dependent nonlinear focusing. (a) Calculated axial distributions of the Kerr-induced refractive index change at different pulse durations. (b) Comparison of simulated and experimentally

determined nonlinear focal shifts as a function of pulse duration. (c), (d) Schematic illustrations of nonlinear focusing and weld formation in the femtosecond and picosecond regimes, respectively.

In summary, this study demonstrates the strong influence of pulse duration on nonlinear focusing and weld formation during ultrafast laser welding of fused silica. The experimentally inferred nonlinear focal shift decreased from 1.75 mm at 262 fs to approximately 0.10 mm at 10 ps, while ICCD imaging showed that differences in plasma-emission position appeared during the initial 2k pulses. The COMSOL simulation incorporating the Kerr effect reproduced the experimental trend, supporting the role of Kerr self-focusing in the observed changes in interaction position and final weld morphology. These findings establish a physical connection between nonlinear beam propagation and weld formation, providing a basis for controlling energy deposition and improving the precision of ultrafast laser joining of transparent materials.

**Funding.** National Key Research and Development Program of China (2023YFB4605503); the State Key Laboratory of Micro-nano Engineering Science (Grant No. MES2027-2).

**Disclosures**. The authors declare no conflicts of interest.

**Data availability.** Data underlying the results presented in this paper are not publicly available at this time but may be obtained from the authors upon reasonable request.